\documentclass[10pt]{article}

\begin{document}

\title{Physics and Five Problems in the \\ Philosophy of Mind}

\author{Stuart Kauffman}

\date{July 12, 2009}

\maketitle

{\par\centering Departments of Biosciences and Physics and Astronomy\\ The University of Calgary \par}
\smallskip 

{\par\centering Signal Processing, Tampere University of Technology \par}
\smallskip 

{\par\centering External Professor, The Santa Fe Institute \par}

\noindent

\begin{abstract}
Since Descartes' dualism, with his res extensa and res cogitans, six fundamental problems in the philosophy and natural history of mind are these: 1. how does mind act on matter? 2. If mind does not act on matter is mind a mere epiphenomenon? 3. What might be the source of free will? 4. What might be the source of a responsible free will? 5. Why might it have been selectively advantageous to evolve consciousness? 6. What "is" consciousness? Many outstanding neuroscientists and philosophers hold to a computational view of mind, driven by the power of computers. Penrose advanced the hypothesis that mind might be quantum coherent and be acted upon by quantum gravity. I agree with Penrose that mind and brain are underwritten by the specific physics of the mind-brain system. I approach the first five of the above six problems based on two physical postulates. First the mind-brain system is a quantum coherent, but reversibly decohering and recohering system. This allows me to answer 1) above, mind does not act causally on brain at all, rather it acausally decohers to classicity (for all practical purposes), hence has consequences for brain and body as matter.  Epiphenomenalism is averted. A quantum mind, because it is acausal on Copenhagen including Born, yields a free will, \textit{but a merely random free will, not a responsible free will}. Second, the most radical part of this article proposes that the quantum classical interface is not always describable by a law: specifically in a special relativity setting, no function, F, maps the present state of the system mind-brain into its future. In its place is a \textit{nonrandom yet lawless process}. I seek in this non-random yet lawless process a source for a responsible free will. Finally, if the quantum-classical boundary can be non-random yet lawless, then no algorithmic simulation of the world or ourselves can calculate the real world, hence the evolutionary selective advantages for evolving consciousness to "know" it may be great. I make no progress on problem 6, the hard problem of qualia.
\end{abstract}

\bigskip 

\pagebreak

\section*{Introduction}

Based on two physical postulates, I approach and hope to resolve five fundamental problems in the philosophy of mind that have plagued us for hundreds of years. Both postulates are testable in principle. If mind depends upon the specific physics of the mind-brains system, mind is, in part, a matter for physicists.

Since Descartes invested the Western mind with res cogitans and res extensa, the seemingly insurmountable philosophic and scientific questions his dualism posed have stalked us. Indeed, a friendly observer of the past 350 years of the philosophy of mind might be forgiven for saying that res cogitans and res extensa, despite all our efforts with Dualism, Materialism, Idealism, and now the Mind Brain Identity Theory, have held us at bay.  I say 'at bay' because it is clear that there is no agreement that we have solved the mighty problems of consciousness and mind, (1,2,3,4).

In the present essay I propose to broach new ground that I hope may help solve five fundamental problems in the philosophy of mind and the evolution of consciousness: 1) How does mind act on matter? 2) If it cannot, is mind a mere epiphenomenon?  3) Whence free will in the face of causal closure in the brain?  More, I hope to make inroads on a fundamental fourth problem, 4) Whence a \textit{responsible} free will.  But there is a further issue I want to discuss: 5) What is the evolutionary usefulness, or selective advantage, of consciousness?  And 6) is there any hope that my tries at 1-5 might shed light on the 'hard problem' of consciousness experiences, of qualia?  The answer to this last question appears, as yet, 'No'. 

All the above questions are deeply familiar, and the subjects of massive efforts by philosophers, (1,2,3), neuroscientists, (5,6), physicists, (7) and others.  I propose to state each of these problems, then tackle them with two physical hypotheses:  First, the mind is a quantum coherent-reversibly decohering-recohering system in the brain.  Thus, following R. Penrose, (8) I believe that consciousness is a problem, at least in part, of the \textit{physical basis} subtending it.  While the arguments I advance differ sharply from those of Penrose, and while he was strongly attacked for suggesting a quantum-consciousness connection, he was courageous, and did much to legitimize the 'C' word in serious scientific discussion.  In this view I sharply differ from those who hope for an emergence of consciousness in a computational mind, (3), whether comprised of chips, neurons, or water buckets. 

The second physical hypothesis is scientifically and philosophically radical. The famous Turing-Church-Deutsch, TCD, principle, (9), states that any physical machine can be simulated to arbitrary accuracy on a universal Turing machine.  This thesis is profoundly related to reductionism and the long held belief, since Descartes, Newton, Einstein, Schrodinger, and Weinberg, (10), that there is a 'Final Theory of Everything' at the base of physics, which explains all that unfolds in the universe by logical entailment.  As we shall see, this view derives from Aristotle's analysis of scientific explanation as deduction: All men are mortal, Socrates is a man, thus, Socrates is a mortal.  As Robert Rosen rightly points out, (11), with Newton, we have eliminated all but one of Aristotle's four causes, formal, final, material and efficient, \textit{retaining only efficient cause in science and mathematized it as deduction.}  Thus, Newton's equations, in differential form, with initial and boundary conditions are 'solved' for the behavior of the system by integration, which is precisely deduction.  This identity of efficient cause with deduction leads directly to the reductionist view held by Weinberg and others. There can be no unentailed events, so emergence is just wrong and there must be a final theory 'down there' from which all derives by entailment. As Weinberg famously says, (10), the explanatory arrows all point downward, from societies to people to organs to cells to biochemistry to chemistry to physics and ultimately to particle physics and General Relativity, or perhaps String Theory, (12).  Turing-Church-Deutsch holds precisely the same view - it is algorithms all the way down so entailment all the way up.  In this view, the universe is a formalizable machine, and we who live in it are TCD machines.  Then we, robot-like can use the inputs from our sensors and calculate all we need to flourish, machines afloat in a machine universe. But then, unfortunately, there is no selective advantage to conscious experience.  Why then, did it evolve?

I will present four lines of reasoning and candidate evidence suggesting that reductionism is very powerful, but powerfully inadequate. I will thus argue that there can be no 'theory of everything' that can explain all that unfolds in the universe by logical entailment, hence that the universe and biosphere in their evolution are not machines, and that the Turing-Church-Deutsch does not hold, (4,13). In such a world, the evolutionary advantages of consciousness may be stunning, \textit{for if we cannot, in principle, calculate the behavior} of a universe, biosphere, animal and human life that is partially lawless yet wonderfully non-random then \textit{there may be a profound advantage to conscious experience}.  It is one way we can understand a partially lawless, non-random, hence non-calculable, universe, biosphere, and free willed human life, and flourish in it.  

I note at the outset that I think the scientific grounds for a quantum mind are presently weak, that it is, at present, an improbable scientific hypothesis, but that it is definitely not ruled out, as we shall see, (4 ).

This article is organized in the following sections. Section 1 discusses dualism and its standard philosophy of mind problems. Section 2 discusses some facts about quantum mechanics needed for my discussion. Section 3 proposes answers to how the mind acts on the brain and mind, that appear to be solved by assuming the mind-brain system is quantum coherent, reversibly decohering to classicity for all practical purposes, FAPP, and returning to a quantum state. Section 4,  I take a first inadequate step towards a free will, it is free but not responsible. Section 5, sketches a physical theory for a quantum decohereing-recohering mind-brain system rather analogous to other theories which, however, do not consider reversible decoherence and recoherence. Section 6 is about possible steps towards a responsible free will. Section 7, I consider several reasons why both reductionism and the Turing-Church-Deutsch principle is inadequate, that open the conceptual door toward partial lawlessness, yet non-random becoming. Other scientists seem to be exploring similar ideas, as I describe, (14,15). I will in Section 8 use lawlessness yet non-randomness as a hoped for avenue to a responsible free will.  In section 9, I discuss why the failure of Turing-Church-Deutsch gives a powerful selective advantage to consciousness. If we and the universe are not TCD, then we cannot compute what will happen. Consciousness seems a sufficient evolutionary solution and is thus selectively advantageous.  In Section 10, I confess that none of the above helps understand the hard problem of qualia in themselves.

I hope the ideas in the article open new philosophic and scientific ground for our considerations.

\section{Dualism and Its Familiar Problems}

Descartes famously supposed mind stuff and material stuff, res cogitans and res extensia.  Res extensia was conceived by Descartes as a machine, driven by Aristotle's efficient causes.  We have held to the efficient cause view of the material world from Descartes to Newton to the present.  As noted it is the logical basis of reductionism and TCD. With Descartes, res cogitans, experience, hovered somehow in our brain/body and somehow nowhere.  The immediate issue that arose for Descartes and all who have followed was: How does mind act on matter?  

The standard form of this problem depends upon causal closure in the material world of efficient causes.  Any event (classical physical event) must have a sufficient classical physical efficient cause. Thus there can be no first cause, and causal closure is required.  Given this view, and the current Mind-Brain Identity theory, the standard concern is that brain events are sufficient causes of later brain events, and there is nothing left over for mind to do to affect the brain.  Worse, there is no obvious way the mind, res cogitans on dualism, 'mind' in a mind-brain identity theory, could manage to act on brain.

You may respond: But on the mind-brain identity theory it is not legitimate to then separate 'mind' from 'brain' and ask how the former acts on the latter. They are identical by hypothesis.  Yes, we can say the words, but we all experience qualia, inter alia with respect to other minds. How can our experiences act on matter on any view at all, including the Mind-Brain identity theory?  As philosopher Michael Silberstein told me: (16), "But it will be said of the mind-brain identity theory: separate the mind aspect from the brain aspect. Now how does the mind act on the brain?"  Then Silberstein repeated the arguments above about causal closure in brain stuff and nothing for mind to do, nor any way for mind to do it to brain and body.

The response to this apparent impass is a retreat to epiphenomenalism: Mind does nothing, in fact, it does not act on brain, it is an epiphenomena of brain.  It is fair to say that no one likes this view.

The third problem, assuming classical matter for the brain and causal closure, is free will.  How can we have free will if the world's becoming, like Newton's laws, are fully deterministic?  Then we cannot have free will in truth. And since all our behaviors are determined, we cannot have morally responsible free will.

One response to this problem now prevalent is an appeal to deterministic chaos in the brain and the thought that only a tiny subset of neurons underpin conscious experience, (5,6).  Then infinitesimal alterations in initial conditions will lie on divergent trajectories with positive Lyapunov exponents, the butterfly will flap energetically, and we will have the illusion of free will.  This view may well be true.  But I want to argue that we do not need it. 

\section{Some Quantum Facts}

We are all familiar with the basics of quantum mechanics, including the familiar Copenhagen interpretation and Born rule under which the time dependent Schrodinger equation propagates a wave of 'possibilities' whose amplitudes, when squared, yield the probabilities of a given quantum degree of freedom being measured in a classical apparatus setting.  This view of quantum mechanics is, as we all know, fully acausal. There is no cause for the radioactive decay event that kill's Schrodinger's cat, just bad luck for the cat.  Beyond Copenhagen, we all know the Bohm and Many World interpretations of quantum mechanics, which few hold in favor.  I will base my discussion on Copenhagen/Born and more recent work.

The central topic of my concern will be 'decoherence' as an account of the emergence of the classical world, or, for purists, the classical world FAPP, for all practical purposes, from the quantum world, (17).  This has been well established in work by Leggett with a quantum system interacting with a quantum oscillator bath, (18). More, decoherence is a well established experimental fact in quantum computing, where it destroys the quantum coherence needed for such computation, (19). 

To be more precise, quantum interference, for example in the two slit experiment, requires that \textit{all the phase information} in the Schrodinger wave, or the sum over all possible histories in Feynman's formulation, arrive at the detector and interact by constructive or destructive interference.  These interactions yield the famous interference effects of quantum mechanics that defy classical explanation.

Decoherence requires considering a quantum or quantum + classical 'system' and its quantum or quantum + classical environment.  The central idea is that quantum phase information is lost from the system to the environment, so the system loses the capacity to exhibit quantum characteristic interference phenomena.  The system can approach classicity FAPP, or for some physicists, a classical mixed state of classical probabilities not quantum probabilities that superimpose.

It is essential to the discussion below that \textit{quantum decoherence, the loss of phase information, is not a causal process} in any sense.  Rather phase information, the heart of quantum possibility waves on Copenhagen and Born, is lost acausally from the system to the environment and typically cannot, in any practical way, be recovered.  

The central implication of this is that decoherence constitutes the passage from the quantum world of possibilities to the actual classical (FAPP) world of physical events, and there is nothing causal in this passage.

Below I will explain possible physical embodiments of my hypothesis that the mind is quantum coherent, but reversibly locally passing to decoherence and recoherence repeatedly.  At this point I will say, however, that such reversible passage from a coherent 'entangled state' to decoherent-classical (FAPP) and back is assured by Shor's theorem that shows for a quantum computer whose quantum degrees of freedom are decohering, that they can be made to recohere to coherence by the injection of information in the now thermodynamically open system, (21).  More, Briegel has published two recent papers showing just such reversible passage from quantum entangled to classical and back repeatedly, (22,23).  

Reversibility of the coherent to decoherent-classical to recoherent quantum states are essential to my hypothesis for I wish the brain to be undergoing such reversible transformations all the time.  If we imagine the coherent spatially extended regions of the brain, as discussed below, to be pink, and the decoherent regions to be increasingly grey as decoherence sets in, I imagine a 3 dimensional volume in the brain where each pixel- volume waxes and wanes pink to grey to pink somewhat like an fMRI temporal image.

\section{How Does the Mind 'Act On' the Brain?} 

This question, which seems deeply difficult to answer for a classical brain, becomes easy to answer in the current framework: \textit{The quantum coherent-decohering-recohering mind does not act on the brain causally at all. Rather, by decohering to classical (FAPP) states, the quantum coherent mind has acausal consequences for the classical "meat" of the brain.}  No causality from res cogitans to res extensa is needed.  Mind acausally has consequences for the classical states of the brain.

We may or may not hold a quantum theory of the mind-brain system to be scientifically plausible at this stage. Nevertheless, I claim that decoherence to classicity FAPP is a substantial candidate to answer our 350 year old question of how the mind 'acts on the brain'. It does not act on the brain causally. It decoheres and this alters the classical state(s) of the brain.

Many, notably Dennett, (2), in Freedom Evolving, would disagree strongly with the need for such a quantum decoherent account. Whatever the merits of Dennett's views, however, they do not vitiate the possibility that a quantum decohering-recohering mind-brain may answer the question of how mind - acausally - has consequences for physical matter.  

Next, how does the either purely quantum mind, or quantum coherent-decohering-recohering mind-brain system act on mind?  A first order answer is Schrodinger's equation itself.  Mind propagates quantum coherent time dependent Schrodinger waves unitarily.  As we will see this is actually not sufficient for a responsible free will, but it is a start, allowing mind to have acausal consequences for the temporal behavior of mind.

With this, we are freed from a retreat into the mind as purely epiphenomenon.  Because we do not have to answer the familiar (classical physics) question of how mind acts efficient causally on brain, the issue of epiphenomenalism does not arise.

\section{A Random Free Will}

We have now a beginning, but inadequate answer to free will.  If we take mind to be quantum coherent, then to decohere to classicity, and take this decoherence to be identical to the standard interpretation of Copenhagen and Born, where the 'collapse of the wave function' occurs upon classical measurement, then the Schrodinger equation gives the \textit{fully acausal fully random probability} of a quantum degree of freedom being measured with a specific value. In the older Copenhagen interpretation, the wave function collapses from all its possible values to a unique classically measured value.

Then since this process is acausal, we do not confront in the quantum realm the issue of classical causal closure, so can have a 'free will'.  This is a start, but not adequate.

The inadequacy of this start of a theory of free will is that this free will is not responsible. Here is the issue: If the mind causally and deterministically determines the brain and our actions, then we do not have free will. Conversely, if the determination of our actions by an acausal quantum mind is \textit{simply randomly probabilistic}, then again, we are not responsible for our actions. We just randomly happen to kill the old man in the wheelchair.  

This is a very deep problem. Attempting to address it will require most of the rest of this article.

\section{A Physical Theory of the Quantum Mind-Brain}

I begin with old and new opinions and facts. Had one asked a physicist twenty or even ten years ago if the human brain could exhibit quantum coherent phenomena, the response, after laughter, would have been that thermalization would have destroyed any vestige of quantum coherence, so the answer was 'No'.

It is therefore astonishing and important that recent results on the chlorophyll molecule, surrounded by its evolved 'antenna protein', has been shown be quantum coherent for almost a nanosecond.  Now the normal time scale for decoherence is on the order of 10 to the -15 second, or a femto-second. Yet these experiments, carried out at 77K, but thought to apply to chlorophyll in plants at ambient temperature, show quantum coherence of an absorbed photon traveling to the reaction center for over 700 femtoseconds, the length of their longest trial, (24). No one expected this.  The authors believe that the quantum coherence increases dramatically the quantum efficiency of the energy gathering process in photo-synthesis.  More, they believe that the evolved antenna protein either suppresses decoherence or induces recoherence.  No one knows at present.  It seems safe to conclude that quantum coherence for on the order of a billionth of a second, a nanosecond, is possible and observerable at body or ambient temperature.  The evolved role of the antenna protein is testable by mutating its sequence.

The time scale of neural activities is a million times slower, in the millisecond range. But it takes light on the order of a millisecond to cross the brain, so if there were a dispersed quantum decohering-recohering mind-brain, reaching the millisecond range is probably within grasp of a quantum theory of the mind-brain system.

The second recent fact, now widely studied by quantum chemists working on proteins, is that quantum coherent electron transfer within and between proteins is possible and almost certainly real.  Because two proteins may coordinate two water molecules, and the electron can pass between the proteins by two pathways, in analogy with the two slit experiment, quantum interference can happen, (25).  

The next fact is that calculations of electrical conductivity between neighboring proteins as a function of the distance between them shows a plateau between 9 and 14 micron separation.  The author, David Beratan (26), believes that this plateau reflects quantum coherent electron transfer at this separation, about right to coordinate a few water molecules between the proteins.  More, quantum coherent electron transfer occurs within proteins.

Now electrons are only one kind of quantum degree of freedom that may transport within and between nearby complex molecules.

The next fact of importance is that the cell is densely crowded with macromolecules.  I do not know the distribution of distances between them, but it is on the order of dozens of angstroms, probably just enough to admit and coordinate the locations of one or more water molecule that then can support quantum coherent electron transport. This is open to investigation experimentally, including the effects of alteration of osmotic effects, swelling or shrinking cells by uptake or removal of water from the cells, on electron transport in cells. Such shrinkage or swelling could surpass the 9-14 angstrom separation needed for quantum coherent electron transport, hence be visible experimentally.

These facts raise the theoretical possibility that a percolating connected web of quantum coherent-decohering-recohering processes could form among and between the rich web of packed molecules in a cell, let alone its membrane surfaces. 
Hammeroff and Penrose (27) have suggested microtubules forming the cytoskeleton of cells as loci of coherent quantum behavior.  Penrose, (8), has suggested that quantum gravity may play a role in the transition to classicity.  Others have suggested a variety of molecular bases for extended molecular structures that might support quantum coherent behavior, (28, 29). As far as I know, I am the only investigator proposing a quantum coherent-decohering-recohering model of the mind brain system, (4). 

In short, we can imagine a physical substrate in cells that could carry a quantum recohereing-decohering, pink and grey, process in cells and between cells.

My own view of the above is that it remains scientifically unlikely, but given the chlorophyll results and quantum chemistry calculations on electron transport, not impossible at all.  

\section{Possible Steps Towards a Responsible Free Will}

I begin with the comment that Aristotle considered four causes, formal, final, material and efficient.  In a simple example of a house, the formal cause of the house is the design of the house, the blueprint. The material causes are the bricks and mortar and beams. The final cause is my responsible free willed decision to build the house. The efficient cause is the actual process of building the house.

Aristotle also offered an account of scientific explanation: The syllogism. All men are mortal. Socrates is a man. Therefore Socrates is a mortal.  Feel the logical force of the conclusion. It underpins our sense that natural law governs the universe rather than compactly describing its regularities.

As noted, Rosen (11) points out that with Newton's laws, initial and boundary conditions and differential equations, Aristotle's maxim for scientific explanation as a deduction snaps into place, for integration of Newton's differential equations constitutes precisely deduction.  More, as Rosen rightly points out, deduction and integration of differential equations becomes the complete mathemization of efficient cause. All other Aristotelian causes were banished from science.  This banishment, this view that all that happens in the universe is to be explained by deduction, lies at the base of our long love of reductionism, Weinberg's dream of a final theory, (10), and current string theory, (12).  If all explanation is by logical entailment, then we reason that there must be a final theory of everything at the base of physics that entails logically all that unfolds in the universe. The Turing-Church-Deutsch is in full harmony with this: It is algorithms computing functions, or deducing from laws, all the way down.  As Descartes hoped, we live in a machine universe and are, res extensa, living machines.  No need for conscious experience, then, just take in data and compute your world and response to it, like a robot seeking an electric plug to get its battery recharged. 

But there are clouds on the reductionist horizon. Physicist Stephen Hawking recently published an article, "Godel and the End of Physics", (30), arguing that it may be the case that no finite set of efficient cause laws will describe the becoming of the universe, including mind.  There may be no finite Theory of Everything.

When a looming crisis such as this arises, it may be wise to question our fundamental assumptions. One of these is our sole reliance in physics on efficient cause laws.

I therefore now want to raise four issues that will take some time. First, should we trust the 350 or 2500 year old belief that all that unfolds in the universe is due to efficient causes?  Thereafter I will raise a second issue: Does the becoming of the biosphere by \textit{Darwinian preadaptations} admit of a sufficient efficient law description?  Third, does the quantum-classical world evolve according to a law?  If not, does an abiotic natural selection and blind final cause play a role in physics given a reversible quantum-classical process? Is this process lawless and random, or lawless and non-random? Fourth, in considering whether the quantum-classical world co-evolves according to law, is there reason outside of a reversible quantum-classical process to doubt that the total process is lawful, yet if not must it be random or can it be non-random?  Later I will try to find an opening for a responsible free will in a possible efficient cause lawless yet non-random evolution of the quantum-classical world.  If true, we will find the possibility of free will but the non-probabilistic character needed, I hope, for a responsible free will. 

\section{Reductionism and the Turing-Church-Deutsch \\ principle are Inadequate}

\subsection{Blind Final Causes}

First I need to discuss the concept of a Darwinian adaptation. Philosopher David Depew recently remarked that an adaptation, once achieved, is a "blind teleology", (31). This is meant in just the same sense as Dawkin's "The Blind Watchmaker", (32). Darwin gave us a startling idea: the appearance of design could arise without a designer.  Thus, Depew envisons no designer, hence "blind teleology". 

Now I ask, can we speak of the opportunity for an adaptation before it occurs?  Consider an organism that is not light sensitive, and an offspring with a red cell that is light sensitive and that constitutes an adaptation. I translate 'A is an opportunity for an adaptation' as 'A is possible. A may or may not occur. If A occurs, it will tend to be selected and go to fixation in the population.' Note that 'tend to go to fixation' is a dispositional term, and is not open to reduction by translation into any set of necessary and sufficient actual physical events. Thus, the achievement of the adaptation in which the red celled organism is selected to fixation arises by a sequence of perfectly good actual efficient causes. But because we cannot prestate necessary and sufficient efficient causes that achieve an adaptation, we cannot have an efficient cause law for how the adaptation will, in fact, be achieved. Thus, the opportunity for the adaptation itself, is not an efficient cause.  It is, instead, a blind final cause.

This is an essential conclusion. I give two examples, one economic, one biological. In the 1980s, in North America, there were lots of television stations, programming, television sets and, of course, couch potatoes. In this economic 'niche', could one hope reasonably to make money inventing the television remote channel changer?  Of course. And money was made on the invention. Now were the television stations, programming, television sets and couch potatoes efficient causes of the invention of the TV remote? No. These conditions are what I will call 'enabling constraints' or enabling conditions, constituting an economic niche into which the TV remote 'fit' and flourished. This is a case, assuming responsible free will (our central issue of course) of Aristotle's final cause, and requires consciousness.

But the same issue arises in the evolution of the biosphere. Species form niches into which other species 'fit'.  New species evolve and create new niches into which yet more new species evolve and fit.  For example rabbits 'make a living' in a 'rabbit niche', even if that niche is hard to define precisely. (So too is the TV remote economic niche hard to define precisely.)

Do we think that the rabbit niche is an efficient cause of the evolution of rabbits?  No!  The rabbit niche is an enabling constraint, or enabling condition, that enabled rabbits to evolve, be selected and flourish.  Here there is no thought of conscious decision as above with the TV remote. Rather, we confront Depew's Blind Teleology and what I want to call 'Blind Final Cause'.  This conclusion is essential, for the rabbit niche did not cause the rabbit by efficient causes - the efficient causes were the actual events that tended, the dispositional term again, to lead to the selection of rabbits that made a living in the rabbit niche.  

But this conclusion means that our reliance on efficient causes as the sole explanation for the unfolding of the universe, or at least the biosphere that is part of the universe, is wrong. Darwin told us so. The selective conditions constitute the enabling conditions which are the Blind Watchmaker.  But in turn, this frees us from the ancient conviction in Western thought that explanation in science can only be in terms of efficient causes - mathematized as deduction, hence reductionism.

One has only to talk to a paleontologist, or better, an historian, to realize that neither seeks to understand the facts of the world, what happened, in terms of laws and deduction.  Realizing the fundamental role of blind final cause in the biosphere, let alone full teleological final cause, assuming responsible free will,  means that there is no Theory of Everything 'down there', nor is all that unfolds in the universe the deductive consequence of such a Final Weinbergian Theory.  It will take a long time, assuming the above is correct, to understand its full implications.

\subsection{Darwinian Preadaptations Cannot be Described by Sufficient Efficient Cause Law, (4, 13)}

Were we to ask Darwin the function of the human heart, he would say it is to pump blood. But we might object that the heart makes heart sounds and moves water in the pericardial sac.  Darwin would say that these are not the function of the heart, pumping blood is, because the heart was selected, so exists as a complex organized structure and functional system in the universe, in order to pump blood. It was of selective advantage.
This is the familiar Darwinian Blind Watchmaker adaptation.

But Darwin also noted that a causal property of an organism of no selective use in the current environment might be of use in a new selective environment, hence be selected. Typically a new function will come to exist. These are called 'exaptations' or Darwinian preadaptations. There is no thought of evolutionary foresight here.

I give two biological examples. Swim bladders are in some fish. The level of air and water in the sac adjusts neutral buoyancy in the water column. Paleontologists believe that swim bladders arose from the lungs of lung fish. Water got into some lungs, now there was a sac with air and water, poised to evolve into a swim bladder. Assume the paleontologists are correct.

Two initial question arise: Did a new function come to exist in the biosphere? Yes, neutral buoyancy in the water column. Did this affect the future evolution of the biosphere? Of course, new species, proteins, niches.

The second example concerns the three middle ear bones of mammals. These evolved from three adjacent jaw bones of an early teleost fish by preadaptations.  This example is important because relational degrees of freedom matter. Were one bone in the skull, one in the spine, and one in the jaw, probably hearing bones would not have evolved.

Now I ask the same two questions. Did a new function come to exist in the biosphere? Yes, hearing. Did this alter the further evolution of the biosphere? Yes, new species, proteins, niches.

Now I come to my critical third question: Do you think you could prestate all possible Darwinian preadaptations for all organisms alive now? Well, we don't know all organisms alive now, so I simplify: Could you prestate all possible preadaptations just for humans?

I've now asked thousands of people. We all agree the answer is 'No'.  Parts of the reasons we seem unable to accomplish this task are these: How would we list all possible selective conditions? How would we know we had completed the list? How would we prestate the one or many relational features of one or several organisms that might become preadaptations?  We all feel utterly stymied. We have no way even to start on this task let alone complete it.

I now introduce the 'Adjacent Possible'. Consider 1000 chemical species in a beaker, and call them the Actual. Let them react by a single reaction step. If new species of molecules are formed, call these the Adjacent Possible of the initial Actual.  This is perfectly defined, given a minimum stable lifetime of a species and standard reaction conditions.

I now point to the Adjacent Possible of the Biosphere. Once there were lung fish, swim bladders were in the adjacent possible of the biosphere. Before there were multicelled organisms, swim bladders were not in the adjacent possible of the biosphere.

Now let us see what we have agreed to, unless you think you really can name all human preadaptations.  What we have agreed to is that we do not know all the possibilities in the adjacent possible of the biosphere!  \textit{Not only do we not know what will happen, we do not even know what can happen.}

The next point concerns probability statements about the evolution of the biosphere by Darwinian preadaptations.  Consider flipping a fair coin 10,000 times. It will come up heads about 5000 times with a binomial distribution. But note that we knew ahead of time all the possible 2 to the 10,000th power outcomes, all heads, all tails and so on. We knew all the possibilities, or the sample space, so could construct a frequency interpretation of probability measure over the space.

But we do not know the set of possible Darwinian preadaptations, the sample space, so cannot construct a probability measure.

Laplace had a different version of probability. If confronted by N doors, behind one of which was a treasure, with no further information, the chance that we pick the right door, he said, is 1/N.  But note that we know N, the number of doors. We do not know N for the biosphere so cannot construct a probability measure for the evolution of the biosphere by Darwinian preadaptations.

Worse, if a natural law is a compact description of the regularities of a process, can we have a sufficient natural law for the emergence of swim bladders?  No. We cannot even state the possibility, let alone the probability, let alone have a description of the regularities of a process. So the becoming of the biosphere by Darwinian preadaptations is partially beyond natural law.

This is a major conclusion: We cannot have sufficient natural law for the evolution of the biosphere by Darwinian preadaptations. Yet such preadaptations are common in the biosphere, let alone the economy, cultural evolution and history.  But if this is true, then there can be no final Theory of Everything from which all that unfolds in the universe is logically entailed. With it, the Turing-Church-Deutsch thesis is very strongly weakened.  No algorithm will simulate the evolution of the biosphere with all the quantum events that did or might have happened. Nor could we confirm which simulation was correct. And by the above argument, the becoming of the biosphere by Darwinian preadaptations is not entailed by any Theory of Everything. 

In its place is a vast creativity in which blind final cause, opportunities for adaptation, and unstatable Darwinian preadaptations partially alter how the biosphere evolves.

\textit{It is critical that we have here a process that is partially lawless, yet also is not random!}  The swim bladder and TV remote succeeded in their contexts. Again, the actual process is not describable by a sufficient natural law, but is also not random. We do not have this concept in our physics or our philosophy.  It bears, I think, on a responsible free will. For we have here a partially lawless but non-random becoming.  We are no longer trapped by deterministic efficient cause law, including deterministic chaos, versus 'merely random probabilistic' views of mind and brain.  The success of the swim bladder and TV remote are not merely random probabilistic chance.  We have, for the first time since Descartes, new freedom of intellectual maneuver.  

What does this process of biological evolution say to entailment from a theory of everything?  No. And what does it say to the TCD thesis?  No. 

\subsection{Reversible Decohrence and Recoherence are Partially Lawless and may be subject to Abiotic Natural Selection Blind Final Cause}

I now discuss a controversial topic. I wish to build my case for a quantum coherent-decohering-recohering responsible free will. I base the transition to classicity on decohrence.  Is it lawful? I argue no, based on a position advocated by Karl Popper in his The Open Universe, (32).  Popper uses his argument to support indeterminacy, hence his Open Universe.  I too argue for an Open Universe elsewhere on Popper's and some of the grounds given above and below, (4,33).

Popper considers the setting of special relativity.  An event A has a past light cone and a future light cone, separated by a zone of possible simultaneity.  B is an event in the future light cone of A, so has its own past light cone that includes all of A's past light cone, but parts of B's past light cone are space-like separated from A's past light cone. It follows that at event A, an observer cannot know the parts of B's past light cone outside of A's light cone. Yet the events in this zone outside of A's past light cone and within B's past light cone can influence the event, B.  \textit{But if an efficient cause law is to be constructable by the observer, then that observer cannot do so prior to event B.  For the situated observer at event A , and before event B, no efficient cause law describes the event B; such a law is unknowable and unconstructable by an observer at A and before B.} 

I now translate this to the decoherence setting. Picture two classical (or quantum) detectors retreating from one another at uniform velocity, the special relativity setting.  Now consider a complex organic molecule in a dense mixture of such molecules.  A pair of entangled particles is emitted by the organic molecules, event A, and fly off, say at the speed of light. Some time later they are detected, one or both, by the two detectors, event B. Then at the event A, (and \textit{before the B event}), of the leaving of the entangled particles from the molecule in question, it is impossible to know what events outside the past light cone of A, but inside the past light cone of B, the detection of one or both entangled particles, may influence the B event. But that B event is instantaneously correlated by EPR and may affect the decohrence of molecule A.  For example the shape of the electron cloud and nuclei positions may be affected, falling into one of two alternative decoherent potential wells.  \textit{Thus, Popper's construction implies that there is no law in detail for decoherence.  There is no efficient cause law, or function, mapping from the space-time region including A and stopping before B, but including the retreating detectors, that maps into the future to B and after event B.  But a law is supposed to be a compact description of the regularities of a process available, like Newton's laws, before, during and after the events unfold.  Then there can be no such law or function.}

But what are the moving detectors?  Special Relativity becomes important at speeds near that of light, but is relevant at any speed of relative motion.  Consider our molecular soup in a cell, crowded with molecules and macromolecules at body temperature, jiggling and folding and unfolding, moving relative to one another as quantum coherent electrons may pass between them.  The relative motions are not constant, but Special Relativity still applies. Each event has a past and future light cone and a zone, small, but finite, because relative motions are small, zone of possible simultaneity.  \textit{No efficient cause function, or law, I claim, describes detailed decoherence in cells.} \textit{No law or function maps the time space region including A and before B occurs, into what happens at B. }If there is a lack of law, an absence of a function, F, that maps from A and its space-time region including the moving detectors, but before B, into a future which includes B, then it appears there can be no theory of everything which entails by deduction beforehand all that happens in the universe, and the TCD thesis is again weakened, and perhaps inadmissible in detail. 

Obviously, this is a new line of thought.  The critical implication that I hope is true is that a quantum decohering-recohering mind-brain identity will propagate trillions of these slightly lawless events. Then, the lawlessness but non-randomness can avalanche so that the longer term behavior of the brain is both lawless yet non-random, and can serve as a basis for a responsible free will, neither deterministic nor 'just random chance'.  I return to this below.

\textit{No Law Describes the Details of Decoherence and Recoherence.}
Both Shor's theorem and Briegel's work imply that recoherence is possible. It may or may not be describable by a law. But if the quantum-classical world is reversible, and decohrence itself is without detailed law available before hand and constructible at A, \textit{then the total process cannot be lawful.}  So the total becoming of the quantum-classical world is beyond sufficient natural law. This seems to imply that no Theory of Everything will describe this becoming, and, as D. d'Lambert, (34), pointed out to me, this seems to imply that the quantum measurement problem is insoluble.
With respect to the quantum mind/brain, this means that there is no efficient cause law for its detailed time evolution.

\textit{Possible Abiotic Natural Selection and Blind Final Cause at the Quantum-Classical Interface.} If quantum to classical is reversible, and if some compositions of classical matter, in their quantum-classical environmental context, are more resistant to returning to the quantum world of mere possibilities, then they will be subjected to an abiotic natural selection in that selective environment, or niche.  \textit{Thus an abiotic natural selection may apply at the quantum-classical interface in appropriate circumstances where the environment has a strong bearing on the decoherence process. It seems plausible that this is true in cells.} If this is correct, the abiotic natural selection, like the Blind Watchmaker, creates environments that are opportunities, blind final causes, for the persistence of any bit of now classical, FAPP, matter.  As that bit of matter evolves by adding or subtracting constituents, fitter variants would be expected to be found.  Like blind final cause in the biosphere, we cannot prestate all the necessary and sufficient conditions of efficient causes th
 at achieves such adaptations.

\subsection{Quantum Decoherence and the Subsequent Behavior of the Quantum-Classical System are Lawless but not Random}

In standard quantum mechanics of, say an electron in a classical box, the physicist uses the classical box as classical boundary conditions and solves for the probability distribution of properties of the electron in the box.  These boundary conditions enter the Hamiltonian of the total system.  

Now I raise a new question: Suppose part of a complex quantum system, say the molecular soup in a cell, decoheres to classicity FAPP and yet this decoherence is somewhat lawless by Popper's arguments above.  \textit{Then if we can ever say of the now classical part of the system that it alters the Hamiltonian of the remaining quantum system, a vexed question, we do not know in detail how the Hamiltonian changes because we do not know in detail how the quantum system decohered, partially lawlessly.} In short, a coherent quantum state propagates unitarily, preserving probability. But the decoherence process is dissipative - phase information is lost, but by Popper above, somewhat lawlessly.  How can we know the detailed classical FAPP state, positions of nuclei, for example, that arise? \textit{We cannot, so cannot recompute the further behavior of the total system.} It is somewhat lawless, (20).

Another way of saying this is that, with decoherence, the system falls to a 'mixed' state where all the probabilities are now classical and drawn from some distribution, say of where the nuclei in the molecule are.  But my claim is that we cannot know that mixed state probability distribution, for we do not know how decoherence happened.  For all we know, the now classical probability distribution of the mixed state could be anything, including sharply peaked over a few alternatives.  Again, the becoming of this system has no efficient cause function or law for its temporal evolution.  Again this casts doubt on the capacity of a Theory of Everything to deduce by entailment all that unfolds in the universe. And it casts doubt on the Turing-Church-Deutsch principle of algorithms all the way down.

Remarkably, Conway and Kochen, in the Free Will Theorem, (14), and the (Strong) Free Will Theorem, (15), on entirely different arguments, reach much the same conclusions. "Some believe that the alternative to determinism is randomness, and go on the say that 'allowing randomness into the world does not really help understand free will" ... "adding randomness also does not explain the quantum mechanical effects described by our theorem. It is precisely the \textit{semi-free} (my emphasis) nature of twinned particles, and more generally of entanglement, that shows that something very different from classical stochasticism is at play here. Although the Free Will Theorem suggests to us that determinism is not a viable option, it nevertheless enables us to agree with Einstein that 'God does not play dice with the Universe'. In the present state of knowledge, it is certainly beyond our capabilities to understand the connection between the free decisions of particles and humans, \textit{but the free will of neither of these is accounted for by mere randomness} (my emphasis) ... The import of the Free Will Theorem is that it is not only current quantum theory, but the world itself that is non-deterministic, so that no future theory can return us to a clockwork universe".  Elsewhere, (14), "Physical theories since Descartes have described the evolution of a state from an initial arbitrary or 'free state' according to laws that are themselves independent of space and time. We call such theories ... Free State Theories".  But "\textit{No free state theory can exactly predict the results of twinned spin one experiments} (my emphasis) ... (\textit{In short, no function, F, maps the current state of the system into its future.} My comment and emphasis.)."We shall see that it follows from the Free State theorem that no free state theory that gives a mechanism for reduction, and \textit{a fortiori}, no hidden variable theory (such as Bohm's) can be made relativistically invariant".  Thus, Conway and Kochen find grounds for \textit{lawlessness - no function maps the present to the future give
n a 'free state', and a non-random 'semi-free' nature of twinned particles.} (My comment and emphasis).  This too casts doubt on a Theory of Everything explaining all that unfolds by deductive entailment and doubt on the Turing-Church-Deutsch principle.  

\section{Responsible Free Will}

The familiar problem of a responsible free will, to state it again is this: If mind or even mind acting on brain, is deterministic, then we have no free will, but perhaps the illusion we do, for example via chaotic dynamics.  Also a classical mind/brain, I note, leaves us with the forever unsolved problem of how mind acts on matter.  A quantum decohering recohereing mind does have consequences for matter, so affords a solution to this 350 year old problem.

Conversely in standard quantum mechanics, on Copenhagen and Born rule, and quantum degrees of freedom, there is only the Schrodinger equation possibility wave, amplitudes squared, and an \textit{acausal fully probabilistic or random chance occurrence of an event, say the radioactive decay that kills Schrodinger's cat,}  as given by that equation. \textit{We obtain a free will but only a random chance free will.  Again there can be no notion of a responsible free will.  Obviously this is insufficient.}

The discussion above has opened new conceptual avenues.  In brief review Blind Final cause, acting as enabling constraints or enabling conditions, can play a non efficient causal role in the evolution of the biosphere, and, if I am right, at the quantum classical reversible boundary with abiotic natural selection.  In short, blind final cause frees us from full reliance on efficient cause and explanation by deduction, yet what happens is both partially lawless, yet non-random. This is surely true for the evolution of the biosphere. There seems no reason not to consider this lawless but non-random evolution of the quantum classical boundary in a system as complex as the brain.  In short, in the case of blind final cause, biological adaptations in general, and economic-technological development, and history, it seems that the process is both partially beyond sufficient efficient cause natural law, yet, importantly, very much context dependent and non-random.  Both the swim bladder and the TV remote were successfully 'selected' in their environment.  We may hope that the same applies to possible abiotic natural selection at the quantum-classical boundary.

But we have an entire second line of consideration, without invoking abiotic natural selection. As just pointed out, the evolution of a quantum-classical reversible system can have no law for its becoming because we do not know how the mixed state of classical probabilities forms its distribution by lawless decoherence to classicity FAPP.  Alternatively, we do not know, after such lawless decoherence, how the Hamiltonian of the entire system changes. (I note that some physicists do not like this step at all, so caution is required.

I comment that there are experimental tests open to test for such lawlessness in two slit-like experiments as the complexity of the entities passed in beams through the slit increase. Anton Zeilinger (35) has shown that Buckmeisterfullerenes interfere. Presumably a stream of rabbits would not.  At the complexity of objects where decoherence sets is, it should be possible to test if that decoherence is fully lawful or yields unstable statistics, perhaps as interference bands start to fade.  In so far as the lawlessness depends upon Special Relativity as in Popper's argument, the speed of relative retreating motion of classical detectors should be positively correlated with signatures of lawlessness. 

More, if lawless decoherence depends upon the complexity of the quantum or quantum plus classical environment, then it is reasonable to assume that decoherence by loss of phase information would occur more readily in a 'dense' and complex quantum environment.  If so, then at that complexity of objects where decoherence sets in, a dense 'beam' of objects would be expected to show more decohrence and less lawfulness, than a rarified beam.  Conceivably evidence for abiotic selection at the quantum-classical boundary could be found.  

What we seek, based on a quantum coherent-decohering-recohering theory of mind and brain, is a use of these ideas to escape the familiar philosophic boxes.  We now have two routes to lawlessness but non-random behavior at the quantum-classical boundary we can consider, either of which may provide the pathway to a responsible free will, rather than a merely 'random' free will: abiotic natural selection and no way to propagate the unknown mixed state distribution.  

What we need is a way for what we can interpret as 'intentions' to shape the decoherence-recoherence process such that the classical happenings are altered as are the quantum aspects of the total system.  One natural role for intentions to play is as enabling constraints shaping the classical matter.  One route is by influencing abiotic natural selection through alterations in the quantum environment that selects for resistance to return to the quantum world. In short, in the context of abiotic natural selection of classical degrees of freedom resistant to return to quantum, the natural assumption is that the 'environment' of the system is itself a complex mixture of dense quantum and classical events which thereby shapes how decoherence to classicity for all practical purposes of a 'system' in that environment occurs, hence what occurs in the actual physical world.  Then this environment shapes the abiotic natural selection which then alters further non-lawful but non-random decoherence and abiotic natural selection.

An alternative pathway rests on lack of lawfulness about the mixed state classical probability distribution.  This can be lawless, because due to lawless decoherence, yet may yield a classical probability distribution with very useful properties for an intending mind. Thus, the probability distribution could become peaked over one or a few alternatives. Mind would have shaped the becoming of the mind-brain quantum decohering-recohering system is a lawless yet non-random way. 

On either of the above accounts above, we seem to have a possibility of a responsible free will.  This account is obviously only schematic at this stage of development.

\section{Why Might Consciousness Be Selectively \\ Advantageous?}

This is a very hard problem. For most examples, an unconscious computerized robot would seem to do as well. Humphries argues that humans are conscious because awareness 'enchants us' so makes us fitter, (36).  It is an enchanting idea and may be right.

The fundamental argument that consciousness is not useful, however, rests on both reductionism and the Turing-Church-Deutsch principle.  According to that principle, we live in a Cartesian machine universe, fully simulable to arbitrary accuracy on a universal Turing machine, and we too are Cartesian machines. Our sensors can pick up the environment and compute what they will, hopefully having been selected to be a useful set of sensors.  But there is no advantage of being aware, of consciousness, of qualia.

\textit{What if TCD is, as I have argued, false?  What if reductionism itself is false, as I have argued. Then the universe is not a deductively entailed unfolding in its becoming, and no universal Turing machine in me can capture or simulate all of that, partially efficient cause lawless but non-random becoming.}

But if this is true, if the universe and we are not TCD, if reductionism is false, and all that happens is not entailed by a final theory down there which is 'simulable' to arbitrary accuracy,  then there may be an enormous advantage to consciousness.  If I am a responsible free willed tiger chasing a responsible free willed gazelle, I can 'see' what the gazelle is choosing freely to do and alter my behavior. But I cannot compute what the free willed gazelle will do.

In short, it seems to me that the putative non-TCD, non-reductionist character of the real universe, other life, animals and us, renders consciousness selectively advantageous.  The degree to which consciousness is selectively advantageous depends upon how far we are from TCD and reductionism in the real universe and our lives.  No one knows, of course, but this seems a fresh start to the problem of why consciousness evolved.

\section{The Hard Problem, Qualia}

Does any of the above help?  I do not think so, at least yet...  It may be that it points to an avenue that might conceivably help someday, but as ever, we have no idea what consciousness 'is'.  I cannot avoid one thought: reductionism is inherently third person, for deduction is mere logical entailment, verifiable by all of us in third person language. And we feel profoundly that 'objective knowledge' must be third person sharable.  Is there some kind of clue here?  All our knowledge of the world is \textit{inherently first person}. Something big seems missing.  As Strawson noted long ago, (37), we can only be in the world as here-now oriented subjects, not objects. How trapped are we by reductionism into a third person 'knowing' view of the world?  \textit{More, being in the world when we do not always know what can happen cannot be a matter only of reason or knowing.} Reason and knowing are then insufficient guides to living our lives.  How are we, then, in the world?  Perhaps if we try to give up third person language as primary, objective, scientific, and focus on being in the world when we cannot know, that may help with the hard problem.

\section*{Conclusions}

I have presented the mind-brain identity theory in the context of two physical theories: first one in which a multiparticle quantum-classical system is capable of decohering \textit{reversibly} to classicity, or classicity for all practical purposes.  This allows mind to have consequences for brain without having to act by efficient cause on brain.  This appears to resolve two outstanding problems in the philosophy of mind that have plagued us since Descartes: how the mind 'acts on'  matter - it does so acausally via decoherence. How does mind act on mind - via the quantum decohering-recohering dynamical behavior of the mind-brain identity system. Second, I have discussed both reductionism and the Turing-Church- Deutsch principle and find both inadequate. Part of this is the inadequacy of a purely efficient cause view of the unfolding of the biosphere and perhaps the quantum-classical boundary, where I suggested in a Special Relativity setting that detailed decoherence is lawless.  No function maps the present slice of space-time into its future.  And I suggested an abiotic natural selection and a complex quantum-classical environment that shapes the decoherence to classicity FAPP, where that environment acts as the intention that non-lawfully but non-randomly shapes the consequences of mind for brain and action.  In the apparent failure of reductionism and TCD, we have new grounds both for a responsible free will and an evolutionary advantage in evolving consciousness.  We are not, on this view, machines, nor is the becoming of the universe a machine open to deductive inference.  All this is quite radical and will need careful scrutiny.  But it seems possible to test for lawlessness at the quantum-classical boundary, and if so, this article is both philosophy and genuine science.

\section*{Acknowledgement }

This work was partially supported by an iCORE grant and a TEKKES grant.


\begin{thebibliography}{99}
\bibitem{20} This view is not accepted by many physicists.
\end{thebibliography}
\end{document}